\documentclass[11pt,fleqn,leqno]{article}
\usepackage{amsmath,amssymb,amsthm}
\usepackage{palatino,pxfonts}
\usepackage{indentfirst,geometry,setspace,fancyhdr,sectsty}
\usepackage{natbib}
\usepackage{relsize,url}
\usepackage{graphicx}
\usepackage{accents}
\usepackage{amsfonts}
\usepackage[bottom]{footmisc}
\usepackage{indentfirst}
\usepackage{caption}
\newcounter{enumi2}

\newtheorem{lemma}[enumi]{Lemma}
\newtheorem{prop}[enumi2]{Proposition}

\begin{document}

\title{Cyclical Behavior of Evolutionary Dynamics in Coordination Games with Changing Payoffs\footnote{I thank Antonio Penta, Dan Quint, Keith Paarporn, and two anonymous referees for their comments and suggestions; Matthew Johnston and Vasily Zemchikhin for helping me with Lyapunov analysis. I am especially grateful to my advison Bill Sandholm (1970--2020) for his time, support, and encouragement during all stages of this project.}}
\author{George Loginov\footnote{Department of Economics, Augustana University, 2001 S Summit Ave, Sioux Falls, SD 57197, USA. Email: gloginov@augie.edu}}
\maketitle
\begin{singlespace}
\begin{abstract}
\noindent The paper presents a model of two-speed evolution in which the payoffs in the population game (or, alternatively, the individual preferences) slowly adjust to changes in the aggregate behavior of the population. The model investigates how, for a population of myopic agents with homogeneous preferences, changes in the environment caused by current aggregate behavior may affect future payoffs and hence alter future behavior. The interaction between the agents is based on a symmetric two-strategy game with positive externalities and negative feedback from aggregate behavior to payoffs, so that at every point in time the population has an incentive to coordinate, whereas over time the more popular strategy becomes less appealing. Under the best response dynamics and the logit dynamics with small noise levels the joint trajectories of preferences and behavior converge to closed orbits around the unique steady state, whereas for large noise levels the steady state of the logit dynamics becomes a sink. Under the replicator dynamics the unique steady state of the system is repelling and the trajectories are unbounded unstable spirals.
\end{abstract}

\section{Introduction}
\noindent Economic models in evolutionary game theory study the dynamics of human behavior in large populations of agents who are assumed to only care about momentary gains and not to be able to change their strategy instantaneously. One standard model consists of a game that the agents are matched to play, a payoff function that describes agents' preferences, and a revision protocol -- a rule according to which the agents receive and act upon opportunities to revise their strategies in the game. Analysis of such models allows one to describe the evolution of the aggregate behavior of the population and to make predictions about the long-run behavior for a given initial population state\footnote{See \citet{Sand} for background on evolutionary games and \citet{Newton} for the survey on the current state of the field.}. 

Since evolutionary models shift the focus of the analysis from the individual to the population level, some standard game theoretic assumptions are weakened as to mitigate the impact of certain (but not necessary all) idiosyncratic characteristics of individuals. It is usually assumed that the agents are myopic and do not take their future payoffs into account in the process of making decisions. Besides that, all interactions are anonymous and players cannot acquire a reputation even if they are matched to play the game repeatedly. In certain cases the agents are not able to observe the population state and thus do not have sufficient information to identify their optimal behavior. In such circumstances not only the outcome of the interaction, but also the trajectory of aggregate behavior and the speed of its evolution start to matter, as the players learn what is optimal during the interaction (rather than before it) through imitation, sampling or similar processes. 

The notion of time takes on importance as well. If it takes significant time for the population to converge to an equilibrium state, it is possible that the interaction can have an impact on the environment in which it takes place. The goal of this paper is to expand the standard evolutionary framework as to account for that possibility. Our approach can be summarized as follows: while the environment determines the direction of evolution of behavior, the latter slowly reshapes the former in response. The changes in the environment are modeled as payoff changes in the underlying game and depend on the aggregate behavior of the population. Thus, we introduce a model of two-speed evolution in which the payoffs in the population game evolve over time, but at a slower pace than that of the agents' behavior\footnote{A number of models introducing environmental feedback into evolutionary games has been developed in biology. See, for instance, \citet{Akcay}, \citet{Weitz}, and \citet{Tilman}.}.

The idea behind this setup is to investigate how the changes in the environment invoked by current aggregate behavior may affect future payoffs and hence alter future behavior of the population. 
In a given population state population dynamics can be either reinforced or slowed down by the changes in the environment. The former is the case of positive feedback, which could be illustrated by the role search engines and recommendation systems play in the Internet search. The prevalence of a popular web page is reinforced by the fact that it appears in the top of the search engine results. The case of negative feedback can be illustrated by file sharing on the internet. A relatively more popular file-sharing platform is more likely to be legally challenged on copyright issues, hence the benefits of coordinating on it attenuate over time. The situations in which the population dynamic is reinforced by the changes in the environment are of less interest than those in which the environment counteracts the behavioral trend, because the presence of positive feedback only affects the speed of change, whereas negative feedback can also alter the direction of change of the aggregate behavior and therefore potentially result in a structurally different outcome. 

The focus of the paper is on the case in which negative intertemporal feedback from aggregate behavior to payoffs is introduced into an interaction based on a symmetric two-strategy game with positive network externalities. Our model assumes a continuum of players with homogenous preferences who are randomly matched to play the game in continuous time. Within the evolutionary framework positive externalities imply that the payoff to a strategy increases in the proportion of the population choosing that strategy. They also guarantee that at any instant at least one symmetric strategy profile is a Nash equilibrium, so that the interaction is either a coordination game or a game with a dominant strategy. 

The feedback from behavior to payoffs is carried out by the payoff adjustment function, on which two assumptions are imposed. First, we assume that for every strategy it is linear and decreasing in the proportion of the population choosing that strategy. While results similar to ours can be obtained in some nonlinear specifications, the linearity assumption helps us maintain tractability. Second, we let all payoffs that correspond to the same strategy grow at the same rate, so that the incentives to coordinate are constant over time. This assumption guarantees that the results are solely due to the interplay between the externalities and that while the payoffs are changing, the nature of the interaction stays the same. Consequently, over time the payoffs to a more popular strategy decrease faster than those to the other strategy. So if strategy A currently yields a better payoff than strategy B, two effects are observed when a fraction of population switches from B to A. First, there is an immediate increase in the payoff to A and a decrease in payoff to B due to coordination. Second, the payoff growth rate of strategy A falls as more agents are utilizing it, whereas the payoff growth rate of B increases. However, as the agents are myopic, they only take the former effect into account when they consider switching strategies. Over time the more popular strategy becomes less appealing, so under negative intertemporal feedback the individual preferences evolve in an equilibrating manner. 

As an illustration of such an interaction one can think of a population of users choosing between two online services, such as file storages, photo sharing websites or social networks. If users create and distribute content, then positive network externalities are always in place and it is beneficial for everyone to coordinate on the same resource. However, over time congestion may emerge even if the number of users of a particular service stays the same, as the amount of data per user would constantly increase\footnote{See \cite{vanderbilt} which documents the steps Facebook had to take in order to keep up with the constantly increasing number of users and amount of user-produced data.}. To deal with this issue, online platforms have to keep increasing per-capita storing capacities and hire more personnel, while simultaneously competing for new users. The users' choices in this case can be treated as myopic since they tend to join the currently best service, rather than anticipate which platform would outperform its competitors in general. In the light of this illustration our model can be viewed as a dynamic extension of \cite{JohKum2010Congestible} which investigates the interplay between congestion and network effects in a static setting. In their model, the equilibrium usage of the platform is derived from the comparison of instantaneous marginal externality effects. We on the other hand emphasize the dynamic nature of congestion, which accumulates over time once the proportion of users of an online service exceeds a particular threshold.

The main consequence of the introduction of payoff adjustments is that the population state alone is not a sufficient statistic for the population dynamics. If payoffs were fixed, the set of states in which a certain strategy is optimal will be fixed, too. With changing payoffs one and the same population state can admit different best responses at different points in time. The key quantity to track is the state at which the agents are indifferent between available strategies\footnote{In coordination games this state coincides with the mixed strategy equilibrium. In games with a dominant strategy this point lies outside the unit interval.}. Since the agents presented with a revision opportunity would likely switch away from the suboptimal strategy, the position of the population state relative to the state of indifference will determine the direction of change in the aggregate behavior. At the same time the aggregate behavior will affect the payoffs and thus adjust the position of the state of indifference. The joint dynamics of strategies and preferences, the former aggregated by the population state and the latter by the state of indifference, is derived in Section \ref{model}. 

Two conditions characterize the steady state of the joint dynamics: the population state must be at rest, and the payoff increments to both strategies must be the same. We consider the best response, logit, and replicator dynamics and demonstrate that in each case there is a unique steady state. The uniqueness is due to the assumption that the payoffs to a strategy decrease the faster the larger the share of population playing that strategy. With respect to the population state the payoff increment function of one strategy will be increasing, whereas for the other strategy it will be decreasing, so there is at most one state in which the increments are the same. The position of that state is independent of the initial conditions of the system and is determined only by the relative speeds of payoff change.

Our analysis demonstrates that cyclical behavior can emerge no matter how slow the payoff adjustment process is. The speed with which payoffs change only affects the sizes of such cycles. We are able to fully characterize the solutions of the best response and replicator dynamics, as well as the logit dynamic with small noise levels. For the logit dynamic with large noise levels we investigate local stability of the steady state. The long run behavior of the population state varies significantly across different dynamics. Under the best response and the logit dynamics with small noise levels all solution trajectories of the system converge to orbits around the steady state, so that in the long run the population state takes all values in a proper subset of the unit interval. Under the replicator dynamic all solution trajectories are unbounded unstable spirals, and the population state visits all points on the unit interval. Finally, under the logit dynamic with large noise levels the steady state is a sink, so the long run prediction for the population state is a single point for all initial conditions in some neighborhood of the steady state. 

The intuition for these results is based on the comparison of the speeds of behavior and preference evolution. In the absence of payoff adjustments it is natural for the whole population to evolve towards coordinating on the same strategy. Yet the smaller the share of players who have not switched to the optimal strategy, the slower the evolution of behavior, as it is less and less likely that the agent who receives a revision opportunity would actually need to switch. Once the payoff adjustment process is introduced, its impact on a strategy is the stronger the larger the share of players who are playing that strategy, so any relative advantage that one strategy has over the other would ultimately be leveled. Therefore the population state would frequently switch the direction of its motion. On the other hand, inertia in the payoff adjustment would prevent the dynamic from converging to the steady state. After the initial advantage of strategy A over strategy B is leveled, for some time strategy A will keep losing its appeal to strategy B because it will still remain more popular and so its payoffs will continue to deteriorate. The only exception is the logit dynamics with large noise levels, in which case payoff difference becomes almost irrelevant and the payoff inertia effect is mitigated.     

The basic idea that the evolutionary process can shape the environment in which it takes place can have two possible interpretations. First, our model can be viewed as a model with state-dependent preferences in the spirit of \cite{Becker96}, in which present choices affect future utility levels whereas the utility function itself is unchanged. Second, it is related to models of two-speed evolution (\cite{PossTSE}, \cite{DekElyYil07}, and, especially, \cite{San01RSC}) based on the 'indirect evolutionary approach' (\cite{GutYaa92}, \cite{Gut95}), in which the evolution of the aggregate behavior of the population shapes the process of 'natural selection' among individual preferences in a heterogenous population. Our model is focused on a different aspect of preference evolution: the fact that one and the same strategy may yield different payoffs as the environment evolves. 

The rest of the paper is organized as follows: In Section \ref{model} we introduce the model of two-speed evolution and derive the joint dynamics of strategies and preferences. In Section \ref{secBR} we characterize the solution for the best response dynamics. To check the robustness of this result, in Section \ref{secLR} we investigate the stability of the steady states under logit and replicator dynamics. Section \ref{c1} concludes. 

\section{The Model} \label{model}

\noindent In this section we describe the strategic interaction between the agents in the population and introduce the rule according to which the agents' preferences evolve. Following that, we derive the joint dynamics of strategies and preferences.

\subsection{The Base Game} \label{basegame}

There is a continuum of agents with homogeneous myopic preferences. The agents are randomly matched to play a symmetric two-strategy game and can only play pure strategies. Time is continuous. The interaction starts at $t=0$, at which point the game is described by a bimatrix in Figure 1 with $a>c$ and $d>b$. Rather than changing their strategies instantaneously, the agents have to make short-term commitments to a strategy. As time passes they randomly receive revision opportunities at a rate that is normalized to 1. 
\begin{figure}[ht]
    \centering
    \includegraphics{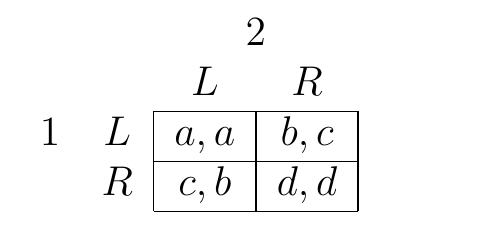}
    \caption{A symmetric two-strategy game.}
    \label{fig:my_label}
\end{figure}

The strategies are denoted $L$ and $R$ for 'Left' and 'Right', correspondingly. We define the population state $x$ to be the proportion of agents who choose $L$. Since the agents are myopic, their preferences are defined as the expected instantaneous payoffs to $L$ and $R$:  
\begin{align*}
f_L(x) = & ax + b(1-x), \text{ and}\\
f_R(x) = & cx + d(1-x) 
\end{align*}
We call the state $y$ which solves $f_L(y)-f_R(y)=0$, namely $y = \frac{d-b}{a-b-c+d}$, the state of indifference. It helps determine which strategy is optimal for each population state: if at time $t$ we have $x_t > y_t$ ($x_t<y_t)$ then Left (Right) is the only best response at $x_t$. In case of equality both strategies are equally profitable. The assumptions on the initial payoffs of the underlying game ($a>c$ and $d>b$) guarantee that the interaction is a coordination game and hence initially the state of indifference lies within the unit interval and coincides with the mixed equilibrium of the game. As the game payoffs change, the type of the interaction might change as well, and the state of indifference might leave the unit interval if one of the strategies becomes dominant. 

\subsection{Evolution of Preferences}

While the standard evolutionary approach postulates that the decisions of individuals in the population are influenced by the environment, the possibility that the environment can be shaped by individual decisions is often overlooked. We model the changes in the environment as payoff changes in the game that the individuals are matched to play. In particular, we focus on the case of negative feedback from aggregate behavior to payoffs, since (in contrast to positive feedback) it can change the direction of the population dynamics. The change in game payoffs will be reflected in the agents' utility function. Therefore we describe that process as evolution of preferences, although no selection among different types of preferences takes place.  

Consider the following scenario: There are two online platforms which provide their services for free to a population of users. The platforms are not perfectly compatible, and users cannot instantly switch from one to the other, so if two users interact they would prefer to use the same platform. However over time the resource that an online service provides depreciates in proportion to the share of population that uses it. As a result, constant utilization of a service adversely affects the benefits of its use, and choice of a platform depends not only on the share of users who are already using it, but also on the relative availability of the resource it provides. 

In accordance with this scenario we let the payoff change be a decreasing function of the corresponding population share. In addition we assume that this function is linear and that the growth rates to payoffs within a strategy are the same, so that the incentives to coordinate remain constant over time, and thus the dynamics of the model can be completely attributed to the interplay between the effects caused by the positive and negative externalities. Let $A$ denote the initial payoff matrix and let $\dot{A}$ be the matrix describing the change in $A$:

\begin{center}
\begin{tabular}{ccc}
$A=\left(\begin{matrix} a & b\\ c & d \end{matrix}\right)$& \text{                    } &$\dot{A}=r \left(\begin{matrix} \hat{x}-(1-k)x & \hat{x}-(1-k)x\\ \hat{x}-k(1-x) & \hat{x}-k(1-x) \end{matrix}\right)$
\end{tabular}
\end{center}

\noindent Parameter $r\ge0$ relates the speed of payoff change (evolution of preferences) to the speed of strategy revision (evolution of behavior)\footnote{While our interest is motivated by cases in which preferences evolve slower than behavior ($r$ is close to 0), our results are qualitatively the same for all positive values of $r$.}. Case $r=0$ corresponds to standard evolutionary models in which payoffs do not change at all. Parameter $k \in (0,1)$ defines the relative depletion rates of the resources. If $k>\frac12$ then payoffs to strategy $Right$ change faster. The constant term $\hat{x}$, which we assume satisfies $0 < \hat{x} < \min \{k, 1-k\}$, guarantees that the payoff to a less popular strategy increases and can be interpreted as the rate at which additional capacity is added by the service. If one plugs in $x=1$ and $x=0$ into $\dot{A}$ one can see that once the aggregate behavior is close to coordination on one of the strategies, the payoffs to that strategy fall whereas the payoffs to the other strategy increase: 

\begin{center}
\begin{tabular}{ccc}
$\dot{A}|_{x=1}=r \left(\begin{matrix} \hat{x}-(1-k)&\hat{x} -(1-k)\\ \hat{x} & \hat{x} \end{matrix}\right)$& \text{                    } &$\dot{A}|_{x=0}=r \left(\begin{matrix} \hat{x} & \hat{x}\\ \hat{x}-k & \hat{x}-k \end{matrix}\right)$
\end{tabular}
\end{center}

\noindent Thus, the negative feedback from aggregate behavior to payoffs affects individual preferences in an equilibrating manner. If Left is currently better than Right, the share of population playing Left will be increasing, so Left will be utilized more and hence it will be losing its advantage over Right. 

\subsection{Joint Dynamics of Strategies and Preferences}
In \ref{basegame} we introduced the indifference state $y=\frac{d-b}{a-b-c+d}$ as the divide between the sets of states in which a certain strategy is optimal. Since it is a function of game payoffs, we can relate its law of motion to the payoff adjustment functions:
\begin{align}
\dot{y} = \frac{d}{dt}\bigg[\frac{d-b}{a-b-c+d}\bigg] = \frac{\dot{d}-\dot{b}}{a-b-c+d}- \frac{(d-b)(\dot{a}-\dot{b}-\dot{c}+\dot{d})}{(a-b-c+d)^2} \label{pffadj}
\end{align}
Following \cite{San01RSC} we call the variable $s=a-b-c+d$ the alignment of the game. It measures the strength of the incentives to coordinate. Indeed, $a-c$ is the gain to coordination if agent's opponent plays Left, and $d-b$ is the gain if the opponent plays Right. The assumption that all payoff parameters of a strategy grow at the same rate guarantees that the incentives to coordinate are constant: $\dot{a}=\dot{b}$ and $\dot{c}=\dot{d}$ imply that $\dot{s}=0$. Hence the second term on the right-hand side of equation \eqref{pffadj} vanishes, and we can express the law of motion of $y$ in terms of population state $x$ as 
\begin{align}
\dot{y} = \frac{\dot{d}-\dot{b}}{s} = \frac{r}{s}\big(\hat{x} - k(1-x) - \hat{x} + (1-k)x\big) = \frac{r}{s} (x - k) \label{pffadj2}
\end{align}

We can now derive the joint dynamics of strategies and preferences. The position of $x$ with respect to the state of indifference $y$ determines the optimal strategy at that population state. At the same time the aggregate behavior affects the payoffs and thus adjusts the position of the state of indifference. If we denote the law of motion of the population state $x$ by $V(x,y)$, the joint dynamics is a system
\begin{align*} 
	\dot{x} &=V(x, y) \\
	\dot{y} &=\frac{r}{s}(x-k)
\end{align*}
with some initial conditions $(x_0, y_0) \in S$ where $S = [0,1] \times \mathbb{R}$ is the set of possible states of the joint dynamics. In the next two sections we examine the behavior of the systems generated by the best response, logit, and replicator dynamics. 

\section{Best Response Dynamics} \label{secBR}
The best response dynamics, introduced in \cite{GilMat91} and \cite{Mat92}, is a deterministic dynamics in which the players use their revision opportunities to switch to the current best response in the population game. Therefore only the players who currently play suboptimal strategies switch strategies. This dynamics requires the population state to be publicly known, so that the agent who receives a revision opportunity can determine which strategy is optimal. When the population state coincides with the state of indifference, there are multiple best responses and hence there can be multiple solution trajectories. 

If preferences do not evolve ($r=0$) this revision rule can generate three types of behavior. If $x_0>y_0$ there will be exponential decay toward the state $x=1$. In other words, if initially enough players play Left, one should expect the whole population to coordinate on that strategy over time. The same reasoning applies to the opposite case: If $x_0<y_0$, the population will move toward the state $x=0$. Initial condition $x_0=y_0$ gives rise to multiple solution trajectories since there are multiple best responses at that state. The system might spend an arbitrary amount of time at the mixed equilibrium before leaving it. 

Since the agents who are not playing the best response switch to it with certainty, at each state except for the state of indifference the speed of the dynamic is determined by the proportion of agents not playing the best response. If $x=y$, both strategies yield the same payoff, hence it is possible that any player would switch. The law of motion of the aggregate behavior can be expressed as 
\begin{align} \label{BR}
\dot{x} = \left\{\begin{tabular}{l c l}
$1-x$     & if & $x>y$ \\
$[-x,1-x]$   & if & $x=y$ \\
$-x$     & if & $x< y$ 
\end{tabular}\right.
\end{align}

The first observation that will help us characterize the global solution of the joint dynamics generated by the best-response protocol is that the solution trajectory from any off-diagonal initial condition intersects the diagonal. Let $D = \{ (x,y) | x \in [0,1] \text{ and } x=y \}$ be the diagonal of the state space $S$. Then as Lemma \ref{lemma1} states, any trajectory that starts in $S\setminus D$ intersects $D$. 
\begin{lemma} \label{lemma1}
Consider the joint dynamics generated by \eqref{pffadj2} and \eqref{BR}. For $r>0$ and any initial condition $(x_0, y_0) \in S\setminus D$ there exists $t^* >0$ such that $x(t^*) =y(t^*)$. 
\end{lemma}
\begin{proof}
Let $S^+ = \{ (x,y) | x \in [0,1] \text{ and } x>y \}$ be the part of the state space below the diagonal and $S^- = \{ (x,y) | x \in [0,1] \text{ and } x<y \}$ be the part above the diagonal. Then the diagonal partitions the state space into three sets: $S=S^+ \cup D \cup S^-$. 

First, consider the initial conditions that lie on the boundary of the state space. If $x_0 = 0$ and $y_0>0$ then according to \eqref{BR} $\dot{x}(x_0, y_0)  = -x_0 = 0$ and according to \eqref{pffadj2} $\dot{y}(x_0, y_0) =\frac{r}{s}(x_0-k) = -\frac{rk}{s} <0$. Hence $x_t$ will remain constant and $y_t$ will be decreasing at a constant rate as long as $x_t < y_t$ until the solution reaches the point $(0,0) \in D$. Similarly, if $x_0 =1$ and $y_0<1$, then $\dot{x}(x_0, y_0) =0$ and $\dot{y}(x_0, y_0)=\frac{r}{s}(1-k) >0$, so $x_t$ will again remain constant and $y_t$ will be increasing at a constant rate until the solution reaches the point $(1,1) \in D$.

If $x_0=0$ and $y_0 <0$, then  $\dot{x}(x_0, y_0) =1$ and $\dot{y}(x_0, y_0) = -\frac{rk}{s} <0$, so the solution immediately escapes into the interior $S^+$. If $x_0=1$ and $y_0>1$, then $\dot{x}(x_0, y_0) =-1$ and $\dot{y}(x_0, y_0)=\frac{r}{s}(1-k) >0$, so the solution immediately escapes into the interior of $S^-$. 

As long as the trajectory of the dynamics remains within the interior of $S^+$ ($S^-$) we can find a closed form solution that characterizes it. Integrating \eqref{BR} yields: 
\begin{align} \label{eq4}
x = \left\{\begin{tabular}{l c l}
$1-(1-x_0)e^{-t}$     & if & $x_0>y_0$ \\
$x_0e^{-t}$     & if & $x_0< y_0$ 
\end{tabular}\right.
\end{align}
If we use \eqref{eq4} to integrate \eqref{pffadj2} we can also describe the trajectory of $y$:
\begin{align} \label{eq5}
y = \left\{\begin{tabular}{l c l}
$\frac{r}{s}(1-k)t+\frac{r}{s}(1-x_0)(e^{-t}-1)+y_0$     & if & $x_0>y_0$ \\
$-\frac{r}{s}kt+\frac{r}{s}x_0(1-e^{-t})+y_0$     & if & $x_0< y_0$ 
\end{tabular}\right.
\end{align}
Next set \eqref{eq4} equal to \eqref{eq5} for the case $x_0 > y_0$ and rearrange it to isolate the exponent:
\begin{align}
& 1-(1-x_0)e^{-t} = \frac{r}{s}(1-k)t+\frac{r}{s}(1-x_0)(e^{-t}-1)+y_0 \nonumber\\
\Rightarrow & 1-y_0 + \frac{r}{s}(1-x_0) - \frac{r}{s}(1-k)t =\big(1-x_0 + \frac{r}{s}(1-x_0)\big)e^{-t}\nonumber\\
\Rightarrow & \frac{1-y_0 + \frac{r}{s}(1-x_0)}{1-x_0 + \frac{r}{s}(1-x_0)}-\frac{\frac{r}{s}(1-k)}{1-x_0 + \frac{r}{s}(1-x_0)}t=e^{-t} \label{eq6}
\end{align}
The left-hand side of \eqref{eq6} is linear in $t$. If we let $A = \frac{1-y_0 + \frac{r}{s}(1-x_0)}{1-x_0 + \frac{r}{s}(1-x_0)}$ and $B =\frac{\frac{r}{s}(1-k)}{1-x_0 + \frac{r}{s}(1-x_0)}$, we can rewrite it as 
\begin{align} \label{eq7}
A - Bt = e^{-t} 
\end{align}
Condition $x_0 > y_0$ guarantees that $A > 1$ and since $x_0 \in [0,1]$ $B$ must be positive. At $t=0$ the value of the left-hand side of \eqref{eq7} exceeds that of the right-hand side ($A>1$), whereas at $t=\frac{A}{B}$ the opposite is true: $0 < e^{-\frac{A}{B}}$. Therefore due to continuity of the functions on both sides of \eqref{eq7} there exists $t^* \in (0, \frac{A}{B})$ such that $x(t^*) =y(t^*)$. 

For $x_0 < y_0$ the resulting equation is similar, 
\begin{align*}
\frac{y_0 + \frac{r}{s}x_0}{x_0 + \frac{r}{s}x_0}-\frac{\frac{r}{s}k}{x_0 + \frac{r}{s}x_0}t=e^{-t},
\end{align*}
and the same reasoning applies. 
\end{proof}
\bigskip

The intuition for the proof is based on the fact that full coordination cannot be attained in finite time if the initial population state $x_0$ is different from 0 or 1. The more players coordinate on the same strategy, the slower the change in the population state (as fewer and fewer players are choosing the suboptimal strategy) and the faster the adjustment of the indifference state, as higher degree of coordination causes more wearing down of the more popular strategy. Conceptually, if $x_0>y_0$ then Left is the only best response, and we should expect $x$ to increase as agents will be switching away from Right. But since Left becomes more utilized, at some point the benefits to its use start to decrease, and $y$ will start increasing as well. But the closer $x$ gets to 1, the faster its speed falls to 0, whereas the speed of $y$ only grows, so ultimately $x$ and $y$ will coincide.   

The proof of Lemma \ref{lemma1} guarantees that we can define a function $\phi : S\setminus D \to [0,1]$ by $\phi(x_0, y_0) = x(t^*)$, mapping any off-diagonal initial condition into the population state at which the trajectory of the solution from that initial condition intersects the diagonal. The next result establishes an important property of this mapping: all initial states below the diagonal are mapped in some neighborhood of 1, whereas all initial states above the diagonal are mapped into some neighborhood of 0. 

\begin{lemma} \label{lemma2}
Consider the joint dynamics generated by \eqref{pffadj2} and \eqref{BR}. There exist $\alpha, \beta \in (0,1)$ such that (i) $\alpha < k < \beta$, (ii) if $(x_0, y_0) \in S^-$ then $\phi(x_0, y_0) \in [0, \alpha)$, and (iii) if $(x_0, y_0) \in S^+$ then $\phi(x_0, y_0) \in (\beta, 1]$.
\end{lemma}
\begin{proof}
First consider $(x_0, y_0) \in S^+$ with $x_0=1$. According to Lemma \ref{lemma1}, trajectories from initial conditions of this type intersect the diagonal at $(1,1)$. \\
Next consider $(x_0, y_0) \in S^+$ with $x_0\in (k,1)$. From \eqref{pffadj2} and \eqref{BR} we establish that 
\begin{align*}
\dot{x}&(x_0, y_0) = 1-x_0> 0, \text{ and}\\
\dot{y}&(x_0, y_0) = \frac{r}{s}(x_0-k)> 0, 
\end{align*}
and hence $ \dot{y}(x_0, y_0) \le \dot{x}(x_0, y_0)$ for $k < x_0 \le \frac{s+rk}{s+r}$. Let $\beta = \frac{s+rk}{s+r}$. We know that $x_0 > y_0$, and as long as $x_0 < \beta$, $x$ grows faster than $y$, so the trajectory of the solution from $(x_0, y_0)$ cannot intersect the diagonal on $[k, \beta]$. But Lemma \ref{lemma1} states that the intersection exists, and since both variables are growing, it must be that $\phi(x_0, y_0) > \beta$. \\
Next consider $(x_0, y_0) \in S^+$ with $x_0 \in (0,k]$. In this case 
\begin{align*}
\dot{x}&(x_0, y_0) = 1-x_0> 0, \text{ and}\\
\dot{y}&(x_0, y_0) = \frac{r}{s}(x_0-k)\le 0, 
\end{align*}
so $x$ is growing while $y$ is declining, therefore the solution trajectory cannot intersect the diagonal on $(0,k]$ and at some point $x$ must exceed $k$. But then the trajectory must go through the region considered in the previous case, so again it must be that $\phi(x_0, y_0) > \beta$.

Finally, let $(x_0, y_0) \in S^+$ with $x_0=0$. Then according to the proof of Lemma \ref{lemma1}, the solution trajectory from such initial conditions immediately escapes into the interior of $S^+$ for which the result holds true. Hence for all $(x_0, y_0) \in S^+$ the solution trajectory intersects the diagonal at some point in $(\beta, 1]$.

For the case of $(x_0, y_0) \in S^-$ we establish that $\alpha = \frac{rk}{s+r}$ and apply the same reasoning to show that $\phi(x_0, y_0) < \alpha$.
\end{proof}
\bigskip

The proof of Lemma \ref{lemma2} is based on a comparison of the signs and absolute values of the speeds of motion of $x$ and $y$. We establish that $x$ increases in all states under the diagonal and decreases above the diagonal, whereas $y$ grows to the right of the line $x=k$ and falls to the left of it. This implies that the trajectories go counterclockwise around the state $(k,k)$. In addition, if the initial condition is below the diagonal, $x$ grows faster than $y$ at all states $(x_0, y_0)$ with $0 < x_0 < \beta$, implying that any trajectory must intersect the diagonal above $\beta$. Similarly, if the initial condition is above the diagonal, the solution trajectory intersects the diagonal below $\alpha$. Possible solution trajectories are illustrated in Figure 2.  
\begin{figure}[ht]
\begin{center}
\includegraphics{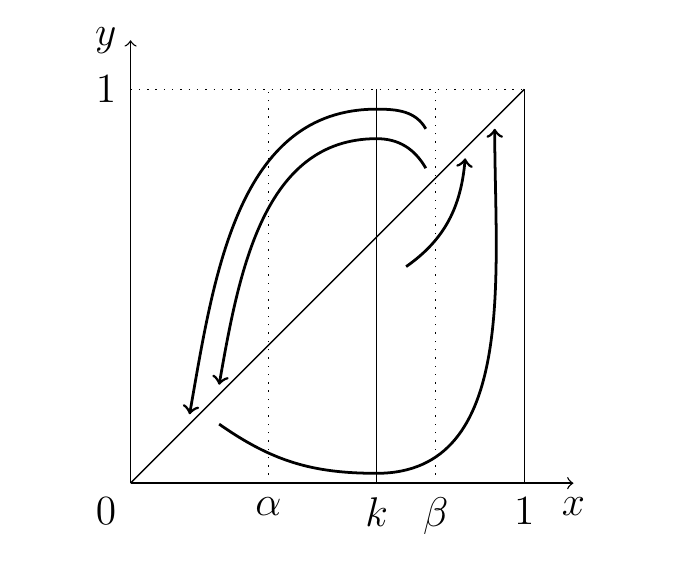}
\end{center}
\caption{Solution trajectories from off-the-diagonal initial conditions.}
\end{figure}

The states on the diagonal require separate consideration, as multiple solution trajectories might emerge when the agents are indifferent between the two strategies. Lemma \ref{lemma3} demonstrates that the dynamics gives rise to multiple solutions only in the vicinity of the state $(k,k)$. On that subset of the diagonal the agents may remain indifferent for some amount of time before the system evolves to some state with a unique best response, whereas on the rest of the diagonal the dynamics admits a unique direction of motion.

\begin{lemma} \label{lemma3}
Consider the joint dynamics generated by \eqref{pffadj2} and \eqref{BR}. Let $(x_0, y_0) \in D$.\\
(i) If $x_0 \in [0, \alpha]$ the solution trajectory immediately escapes the diagonal in the direction of $S^+$. If $x_0 \in [\beta, 1]$ the solution trajectory immediately escapes the diagonal in the direction of $S^-$.\\
(ii) If $x_0 \in (\alpha, k) \cup (k, \beta)$, solution trajectories from $(x_0, y_0)$ can remain on the diagonal for some amount of time before leaving it in either direction.\\
(iii) If $x_0 = k$, the system can spend an arbitrary amount of time on the diagonal before leaving it in any direction. 
\end{lemma}
\begin{proof}
(i) For the solution trajectory to remain on the diagonal during a time interval $[0, T]$ the rates of change of $x$ and $y$ must coincide almost everywhere on this interval. Using \eqref{pffadj2} and \eqref{BR} we can write the condition $\dot{x}(x_0, y_0) = \dot{y}(x_0, y_0)$ as 
\begin{align} \label{ineq8}
-x_0 \le \frac{r}{s}(x_0-k) \le 1-x_0 \Leftrightarrow \alpha \le x_0 \le \beta
\end{align}
Therefore if $0 \le x_0 < \alpha$ or $\beta < x_0 \le 1$ the solution trajectory must immediately escape the diagonal. Since on the diagonal the rate of change of $x$ can be positive or negative, it can in principle escape into either $S^+$ or $S^-$. However, if $0 \le x_0 < \alpha$, the trajectory can only escape into $S^+$, because for any point $(x',y') \in S^-$ with $x'<\alpha$ the rates of change of both $x$ and $y$ will be negative with $\frac{ \dot{y}(x',y')}{\dot{x}(x',y')}>1$, so the vector field at $(x',y')$ will be directed towards the diagonal, whereas for $(x'',y'') \in S^+$ with $x''<\alpha$ the rate of change will be positive for $x$ and negative for $y$, so the vector field at $(x'', y'')$ will be directed away from the diagonal with $\frac{ \dot{y}(x'',y'')}{\dot{x}(x'',y'')}<0$. Since the dynamics is sufficiently smooth in $S^+$, the direction in which a solution through state $(x_0, y_0)$ escapes the diagonal is unique. 

In the same fashion, if $\beta < x_0 \le 1$, the trajectory cannot escape into $S^+$ because both $x$ and $y$ must grow in that region, and $y$ must grow faster than $x$. See (ii) for cases $x_0 = \alpha$ and $x_0=\beta$.\smallskip

(ii) Generalizing the argument from (i) for $(x,y) \in S^-$ we can claim that $\frac{ \dot{y}(x,y)}{\dot{x}(x,y)} < 1$ for $x \in (\alpha, 1]$, whereas for $(x,y) \in S^+$ $\frac{ \dot{y}(x,y)}{\dot{x}(x,y)} < 1$ if $x \in [0, \beta)$. Therefore for $\alpha < x_0 < \beta$ the solution trajectory can escape the diagonal in either direction, as the vector field is pointing 'the right way' in the neighborhood of that part of the diagonal. 

Since all $x_0 \in (\alpha, \beta)$ satisfy  \eqref{ineq8}, the solution trajectory passing through these points on the diagonal does not need to leave the diagonal immediately. For all initial conditions $(x_0, y_0) \in D$ with $x_0 \in (\alpha, k)$ the rate of change of $y$ is negative: $\frac{r}{s}(x_0-k) < 0$, so in order for the solution to remain on the diagonal, $x$ and $y$ must fall at the same rate at almost all times. In this case the solution will be moving toward the point $(\alpha, \alpha)$. Since on the interval $(\alpha, x_0)$ the rate of change is bounded away from 0, the point $(\alpha, \alpha)$ can be reached in finite time for any $x_0 \in (\alpha, k)$. 

If $x_0= \alpha$ the \eqref{ineq8} is still satisfied, but since $\dot{y}(\alpha, \alpha) < 0$, the solution cannot move up the diagonal. It cannot go down either because in the region $[0, \alpha)$ it must immediately leave the diagonal. Since for  $(x,y) \in S^-$ with $x= \alpha$ the slope $\frac{\dot{y}(x,y)}{\dot{x}(x,y)}$ equals 1, the solution cannot escape into $S^-$. Therefore the only direction of escape is $S^+$ in which $\frac{ \dot{y}(x,y)}{\dot{x}(x,y)} <0$ in the neighborhood of $(\alpha, \alpha)$. 

For all initial conditions $(x_0, y_0) \in D$ with $x_0 \in (k, \beta)$ the rate of change of $y$ is positive, so the solution has to move up the diagonal if it is to remain on it. Similarly to $x_0= \alpha$, at $x_0= \beta$ the solution can only escape into $S^-$. \smallskip

(iii) Both $\dot{x}$ and $\dot{y}$ are at rest at the point $(k,k)$, so the joint dynamics can spend an arbitrary time at that state before leaving it in any direction. Ultimately, random fluctuations in the rate of change of $x$ will force the solution trajectory out of that state, but there is no definite moment when that happens. 
\end{proof}

Figure 3 combines the conclusions of Lemmas \ref{lemma2} and \ref{lemma3}. Solution trajectories passing through the diagonal at states $x \in [0, \alpha]$ and $x \in [\beta, 1]$ escape the diagonal immediately into $S^+$ and $S^-$, correspondingly. For $x \in (\alpha, k)$ solutions can move along the diagonal toward the point $(\alpha, \alpha)$ prior to escaping the diagonal into either $S^-$ or $S^+$. For $x \in (k, \beta)$ solutions can move along the diagonal toward the point $(\beta, \beta)$ or escape in either direction. Solutions from the state $(k, k)$ can spend arbitrary time at rest before escaping in any on- or off-diagonal direction. If a solution trajectory escapes the diagonal into $S^+$, it must eventually return to the diagonal between states $\beta$ and 1. If it escapes into $S^-$, the next intersection with the diagonal occurs on the set $[0, \alpha)$. 
\begin{figure}[ht]
\begin{center}
\includegraphics[scale=1]{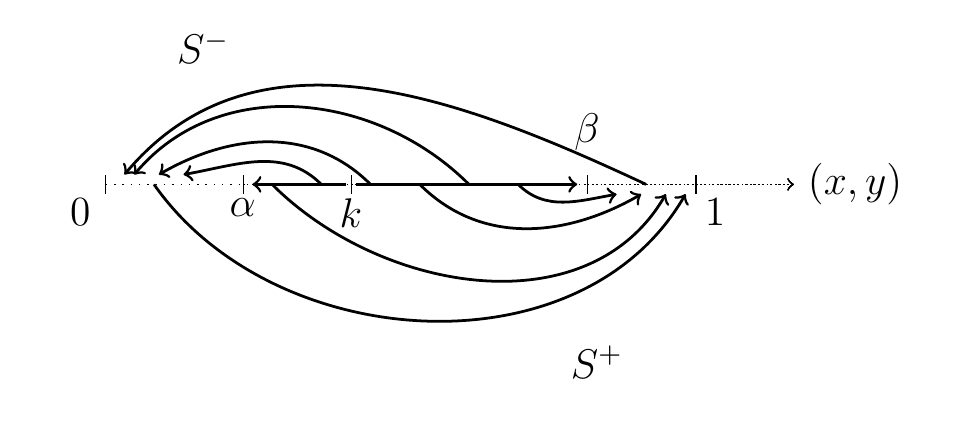}
\end{center}
\caption{Solution trajectories from the initial conditions on the diagonal.}
\end{figure}

We have thus considered all possible initial conditions of the system and therefore can conclude (with minor qualifications) that the joint dynamics of strategies and preferences exhibits cyclical behavior around the state $(k, k)$. Due to the fact that the best response is not unique at the states on the diagonal, certain initial conditions assume multiple solution trajectories, however, all of them exhibit the same limiting behavior. The key feature of the system is that it will visit a certain subset of the diagonal infinitely often whereas certain states in the neighborhood of the steady state will never be visited once abandoned. To see this note that once a solution trajectory has left the diagonal for the first time, it can only return to the diagonal at states with $x \in [0, \alpha) \cup (\beta, 1]$. This helps explain why the only steady state is always repelling. Proposition \ref{T1} provides the complete description of the behavior of the dynamics.  

\begin{prop} \label{T1}
Consider the joint dynamics generated by \eqref{pffadj2} and \eqref{BR}. \\
(i) The only steady of the system is $(k, k)$. \\
(ii) The steady state is repelling, and any solution trajectory from other initial conditions converges to a closed orbit around it.  
\end{prop}
\begin{proof}
(i) One can verify that the only state that satisfies the condition 
\begin{align*}
\dot{x}&(x, y) = 0, \text{ and} \\
\dot{y}&(x,y) = 0 
\end{align*}
is the state $(x,y) = (k,k)$. Hence the steady state of the joint dynamics exists and is unique. \smallskip

(ii) Lemma \ref{lemma1} demonstrates that any solution trajectory from an off-diagonal initial condition intersects the diagonal in finite time. Lemma \ref{lemma2} guarantees that that intersection takes place at a state with $x \in [0, \alpha) \cup (\beta, 1]$. Conversely, Lemma \ref{lemma3} states that any trajectory that goes through a state on the diagonal other than $(k,k)$ leaves the diagonal in finite time. Therefore any solution trajectory would go through and leave the diagonal infinitely many times. 

Next denote the time intervals during which the trajectory is off the diagonal the iterations of the trajectory. Namely, an iteration is an interval $T=(t_1, t_2)$ such that $x(t_1) = y(t_1)$, $x(t_2) = y(t_2)$ and for all $t \in (t_1, t_2)$ we have $x(t) \ne y(t)$. Clearly, no two iterations intersect, and each solution trajectory contains countably many iterations, because it can only spend a finite time off-diagonal before returning to the diagonal and vice versa. Our claim is that each solution trajectory can be described by a sequence $\{x(t_n)\}$ with $n \in \mathbb{N}$, such that each interval $(t_n, t_{n+1})$ is an iteration, so after the first iteration any trajectory doesn't stay on the diagonal for more than a moment. Indeed, if the initial condition is not on the diagonal, then by Lemma \ref{lemma2} the population state $x$ at which the solution trajectory intersects the diagonal belongs to the set $[0, \alpha) \cup (\beta, 1]$, but by Lemma \ref{lemma3} solution trajectories that pass through states in that set must immediately leave the diagonal. If on the other hand the initial condition lies on the diagonal but does not coincide with the steady state, a trajectory might spend only a finite time on the diagonal before leaving it (Lemma \ref{lemma3}), and once it has left it, it can only intersect the diagonal at states in the set $[0, \alpha) \cup (\beta, 1]$, for which the previous argument applies. Moreover, the fact that the dynamics is continuously differentiable in both $S^-$ and $S^+$ guarantees that the direction of motion through all $x \in [0, \alpha) \cup (\beta, 1]$ is unique despite that on the diagonal the dynamics can admit multiple values.  

The next observation is that if $x(t_n) \in [0, \alpha)$ then $x(t_{n+1}) \in (\beta, 1]$ and vice versa. If $x(t_n) \in [0, \alpha)$ the trajectory can only escape into $S^+$ (by Lemma \ref{lemma3}), but then $x(t_{n+1}) \in (\beta, 1]$ (by Lemma \ref{lemma2}). If $x(t_n) \in (\beta, 1]$, the trajectory escapes into $S^-$ and $x(t_{n+1}) \in [0, \alpha)$. Thus after the first iteration any solution trajectory exhibits cyclical behavior in the sense that it sequentially goes through set $S^+$, intersects the diagonal at a state in $(\beta, 1]$, goes through set $S^-$, and intersects the diagonal at a state in $[0, \alpha)$ to start over again. 

Finally, to show that every solution trajectory converges to an orbit, we note that the subsequences $\{x(t_{2n})\}$ and $\{x(t_{2n+1})\}$ with $n \in \mathbb{N}$ are monotonic. This is due to the fact that for each off-diagonal state there is a unique solution trajectory that passes through it. Let $x(t_n), x(t_{n+2}) \in (\beta, 1]$ and assume that $x(t_n) > x(t_{n+2})$. Then as the parts of the solution trajectory corresponding to the iterations $(t_n, t_{n+1})$ and $(t_{n+2}, t_{n+3})$ cannot intersect, it must be that $x(t_{n+1}) < x(t_{n+3}) < \alpha$. Applying the same logic to iterations $(t_{n+1}, t_{n+2})$ and $(t_{n+3}, t_{n+4})$ we conclude that $x(t_{n+2}) > x(t_{n+4}) > \beta$. So if the subsequence with elements in $(\beta, 1]$ is decreasing, the corresponding subsequence with elements in $[0,\alpha)$ must be increasing. Conversely, if the subsequence in $(\beta, 1]$ is increasing, the one in $[0,\alpha)$ must be decreasing. Since both subsequences are bounded and monotonic, they converge. Therefore the full solution converges to a closed orbit around the steady state. 
\end{proof}

\begin{figure}[ht]
\centering
\includegraphics[scale=0.5]{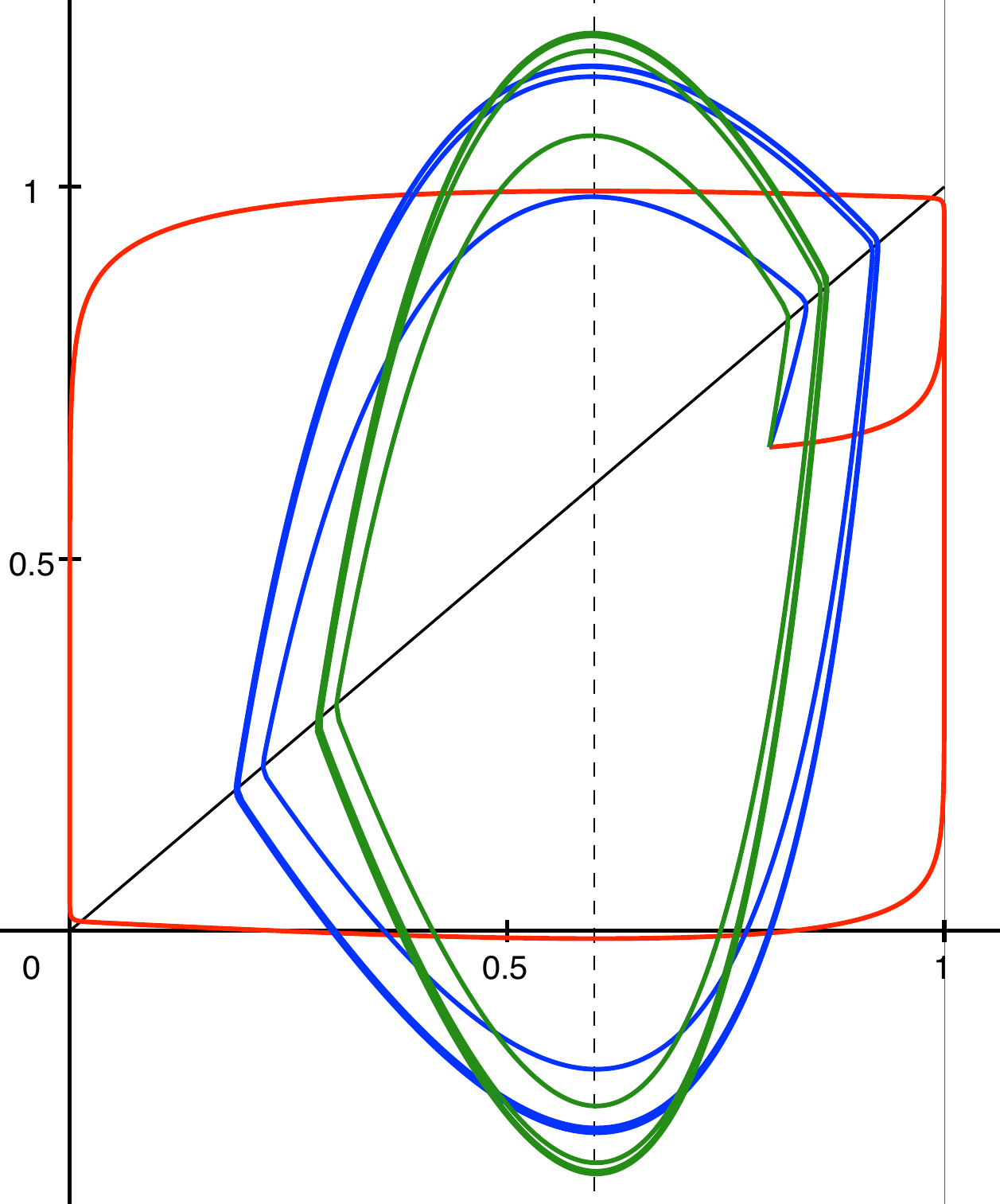}
\caption{Solution trajectories for the BRD with parameters $k=0.6$ and $s=1$ from the initial condition $(x_0,y_0)=(0.8,0.65)$: $r=0.1$ in red, $r=3.5$ in blue, $r=7$ in green.}
\end{figure}

We can summarize the joint behavior of the population state $x$ and the state of indifference $y$ using the following intuition. Assume that initially $x_0 > y_0 > k$ as in the example in Figure 4. Then Left is the best response, and $x$ starts growing as agents switch away from Right. At the same time, as $y_0 > k$, $y$ is growing, too, whereas the payoffs to Left start decreasing. The closer $x$ is to 1, the more slowly it grows, whereas $y$ accelerates, so at some point $y$ will coincide with $x$. At this moment we will observe a switch: the agents will be indifferent between the two strategies while $y$ is still growing, so that at the next moment the situation will be described by $x_t < y_t$. Then $x$ starts decreasing since Right is the new best response. But as long as $x_t > k$, $y$ will continue to grow, so for a while $x$ and $y$ will be moving in opposite directions. As $x$ falls to $k$, $y$ starts to fall, too, and it will overtake $x$ at some state below $k$. So whenever values of $x$ and $y$ coincide, $x$ changes its direction of motion until the next time $y$ overtakes it. After $x$ changes the direction, the variables continue in different directions until $x$ equals $k$, at which point $y$ changes its direction, too. Thus we observe inertia in the behavior of $y$, which in this case prevents the system from converging to the steady state. 

Figure 4 also illustrates the role that the speed of payoff change (parameter $r$) plays in the results. The population state $x$ in the long run will take all values from an interval that is at least $(\beta-\alpha)$ long. Since $\alpha=\frac{rk}{s+r}$ and $\beta=\frac{s+rk}{s+r}$, we have $(\beta-\alpha)=\frac{s}{s+r} \to 1$  as $r\to 0$. So the smaller the speed of the payoff change, the longer it will take for $y$ to catch up with $x$, and hence the wider the interval that contains the values of $x$ in the long run. 

\section{Logit and Replicator Dynamics} \label{secLR}

In this section we derive the equations that describe the law of motion for logit and replicator dynamics. We show that under the logit dynamic with small noise levels the solution trajectories converge to closed orbits around the steady state, thus exhibiting the same behavior as under the best response dynamic. If the noise level is large the unique steady state of the logit dynamic becomes a sink. Under the replicator dynamic the steady state is always repelling, and the solutions form unbounded unstable spirals.  

\subsection{Logit Dynamics}

The logit dynamics, introduced in \cite{Blu93} and \cite{FudLev98}, is an example of a perturbed best response dynamics. In the logit case the switch rate which determines the probability that an agent who receives a revision opportunity would switch from strategy Right to strategy Left is an exponential function of the payoff to strategy Left and has the form $\exp(\eta^{-1} f_R(x))$, where $\eta >0$ is the noise level. In a two strategy game the probability of choosing strategy Left at the population state $x$ can be expressed in terms of the difference in payoffs:
\[
\text{P(choose Left)} =\frac{\exp({\eta^{-1} f_L(x)})}{\exp({\eta^{-1} f_L(x)})+\exp(\eta^{-1} f_R(x))} = \frac{\exp({\eta^{-1} (f_L(x)-f_R(x))})}{\exp({\eta^{-1} (f_L(x)-f_R(x))})+1} 
\]
Thus upon receiving a revision opportunity the agent is most likely to switch to the current best response, however the higher the noise level the higher the chance he would choose some other strategy 'by mistake'. If Left is the only best response, then the probability it will be chosen tends to 1 as $\eta$ approaches 0. If both strategies are best responses, then the likelihood of choosing either of them is $\frac12$. Given the switch probabilities we can derive the mean dynamic by calculating the increment in the number of agents choosing to play Left: 
\begin{align*}
\dot{x} =& \text{P(choose Left }| \text{ currently Right)} - \text{P(choose Right }| \text{ currently Left)}=\\
=& (1-x) \frac{\exp({\eta^{-1} (f_L(x)-f_R(x))})}{\exp({\eta^{-1} (f_L(x)-f_R(x))})+1} -x (1-  \frac{\exp({\eta^{-1} (f_L(x)-f_R(x))})}{\exp({\eta^{-1} (f_L(x)-f_R(x))})+1}) = \\
=&  \frac{\exp({\eta^{-1} (f_L(x)-f_R(x))})}{\exp({\eta^{-1} (f_L(x)-f_R(x))})+1}  - x 
\end{align*}
Although an individual's choice is stochastic, the average behavior of the process can be well approximated by its mean dynamic, since the idiosyncratic noise is averaged away when the population size is large (\cite{BenWei03}). 

Our next step is to express the difference in payoffs in terms of the population state $x$, the state of indifference $y$, and the alignment $s$: 
\begin{align} \label{eq9}
f_L(x)-f_R(x) = ax +b(1-x) -cx - d(1-x) = sx - (d-b) = s(x - y)
\end{align}
Using \eqref{eq9} we can derive the law of motion of the joint dynamics for the logit case: 
\begin{align}
\dot{x} &=\frac{\exp({\eta^{-1} s(x-y)})}{\exp({\eta^{-1} s(x-y)})+1}-x \label{eq10}\\
\dot{y} &=\frac{r}{s}(x-k) \tag{2}
\end{align}
We proceed by showing that this system admits a unique steady state which is repelling when the noise level $\eta$ is small and attracting when it is large. Moreover, when $\eta$ is small the solution trajectory from any initial condition other than the steady state converges to a closed orbit around it. However, we don't establish that this orbit is unique for all initial conditions.  

\begin{prop} \label{T2}
Consider the joint dynamics generated by \eqref{eq10} and \eqref{pffadj2}. For any $r >0$ \\
(i) state $(x^*, y^*)= (k, k -\frac{\eta}{s} \log{\frac{k}{1-k}})$ is the only steady state,\\
(ii) there exists $\eta^*>0$ such that for all $\eta \in (0,\eta^*)$ state $(x^*, y^*)$ is repelling and for $\eta > \eta^*$ state $(x^*, y^*)$ is a sink, \\
(iii) if $\eta \in (0,\eta^*)$ and $(x_0, y_0) \ne (x^*, y^*)$ the solution trajectory from $(x_0, y_0)$ converges to a closed orbit around $(x^*, y^*)$. 
\end{prop}
\begin{proof}
(i) The steady states are the rest points of the dynamic, so we set \eqref{eq10} and \eqref{pffadj2} equal to 0  
\begin{align*}
\frac{\exp({\eta^{-1} s(x-y)})}{\exp({\eta^{-1} s(x-y)})+1}-x=0,\\
\frac{r}{s}(x-k)=0
\end{align*}
The second equation implies that $y$ is only at rest when $x=k$. Then the first equation can be rewritten as
\begin{align*}
\frac{\exp({\eta^{-1} s(k-y)})}{\exp({\eta^{-1} s(k-y)})+1}=k
\end{align*}
so that $\exp({\eta^{-1} s(k-y)}) = \frac{k}{1-k}$ and $y= k - \frac{\eta}{s} \log{\frac{k}{1-k}}$. Therefore the only steady state is $(k, k-\frac{\eta}{s} \log{\frac{k}{1-k}})$.

(ii) To investigate stability of the solution, we linearize the system $F$ generated by \eqref{eq10} and \eqref{pffadj2} around the steady state, letting $R=\exp({\eta^{-1} s(x-y)})$ to simplify notation. The Jacobian of $F$ is  
\begin{center}
$DF(x,y)=\left(\begin{matrix} \frac{sR}{\eta(R+1)^2}-1 & -\frac{sR}{\eta(R+1)^2}\\ \frac{r}{s} & 0 \end{matrix}\right)$  
\end{center}
At the steady state $(x^*,y^*)$ it becomes
\begin{center}
$DF(x^*,y^*)=\left(\begin{matrix} \frac{s}{\eta}k(1-k)-1 & -\frac{s}{\eta}k(1-k)\\ \frac{r}{s} & 0 \end{matrix}\right)$  
\end{center} 
The characteristic polynomial is
\begin{align*}
\lambda^2-\big(\frac{s}{\eta}k(1-k)-1\big)\lambda+ \frac{r}{\eta}k(1-k)=0
\end{align*}
Since $\frac{r}{\eta}k(1-k) > 0$ for $k \in (0,1)$, $r>0$ and $\eta >0$, the roots of the polynomial must have the same sign if they are real. Then the real roots will both be positive if $\frac{s}{\eta}k(1-k)-1>0$. If the roots are complex, the same condition guarantees that their real parts are positive. Therefore the system will be unstable as long as 
\begin{align*}
\frac{s}{\eta}k(1-k)-1> 0 \Rightarrow \eta < sk(1-k)
\end{align*}
and hence $\eta^* =sk(1-k)$.

For $\eta > \eta^*$ the roots of the polynomial are either real and negative or complex with negative real parts, so the steady state is a sink. 

(iii) If $\eta \in (0,\eta^*)$ the steady state is repelling, so all solutions from the initial conditions $(x_0, y_0) \ne (x^*, y^*)$ are bounded away from $(x^*, y^*)$. We will construct a closed and bounded set which will contain no steady states and will be positive invariant for the dynamic system. Solution trajectories from all initial conditions in that set (which we will call a trapping region) will be enclosed in it, and thus in the absence of steady states each of them would converge to a closed orbit. 

\begin{figure}[h]
\centering
\includegraphics[scale=0.25]{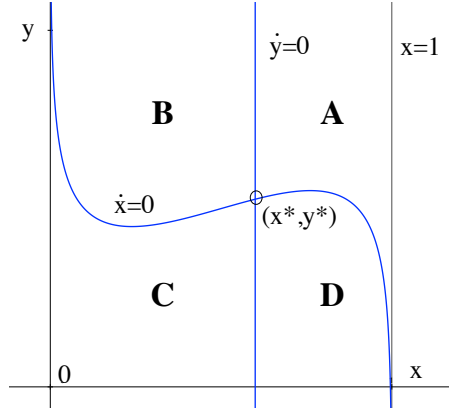}
\caption{The partition of the state space with respect to the signs of $\dot{x}$ and $\dot{y}$.}
\end{figure}

First, observe that isoclines $\dot{x} = p$ are defined by $y_p(x) = x - \frac{\eta}{s} \log (\frac{x+p}{1-x-p})$ on the domain $(-p, 1-p) \cap [0,1]$ with $\dot{x} < p$ whenever $y > y_p(x)$. The curve $y_p$ has the following properties: $y_p \to \infty$ as $x \to -p+0$, $y_p \to -\infty$ as $x \to 1-p-0$, and $y'_p(x) = 1- \frac{\eta}{s(x+p)(1-x-p)}$. The critical point of $y'_p(x)$ is $x=\frac{1}{2}-p$. Therefore $y_p(x)$ is either decreasing on the whole domain when $y'_p(\frac{1}{2}-p) \le 0$ or increasing in some neighborhood of $\frac12 - p$ when $y'_p(\frac{1}{2}-p) > 0$ and decreasing elsewhere. The condition $y'_p(\frac{1}{2}-p) \le 0$ is equivalent to $\eta \ge \frac{s}{4}$. But since we only consider $\eta \in (0,\eta^*)$, it must be that $\eta <sk(1-k) \le \frac{s}{4}$, so the the isocline $\dot{x}=p$ must always have an increasing segment. 

Next observe that the nullcline $\dot{y} = 0$ is defined by $x=k$, so the trajectories through states with $x \in [k, 1]$ cannot flow down, whereas the trajectories through states with $x \in [0, k]$ cannot flow up. The nullclines $\dot{x} = 0$ and $\dot{y} = 0$ split the state space into four regions: $A =\{ (x, y) | \dot{x}<0, \dot{y}>0\}$, $B =\{ (x, y) | \dot{x}<0, \dot{y}<0\}$, $C =\{ (x, y) | \dot{x}>0, \dot{y}<0\}$, and $D =\{ (x, y) | \dot{x}>0, \dot{y}>0\}$ (see Figure 5).  

The trajectories passing through states in regions $B$ and $D$ must flow southwest escaping into $C$ and northeast escaping into $A$, correspondingly. Now consider an initial condition $(x_0, y_0) \in A$. Let $\dot{x}(x_0, y_0) =p< 0$, then the initial condition belongs to the isocline $\dot{x}=p$. Pick $y_1 > \max_{x \in [k, 1]} y_p(x)$ and consider the trajectory passing through $(1, y_1)$ (dashed curve from $(1, y_1)$ to $(x_3, y_3)$ in Figure 6).  

\begin{figure}[ht]
\centering
\includegraphics[scale=0.4]{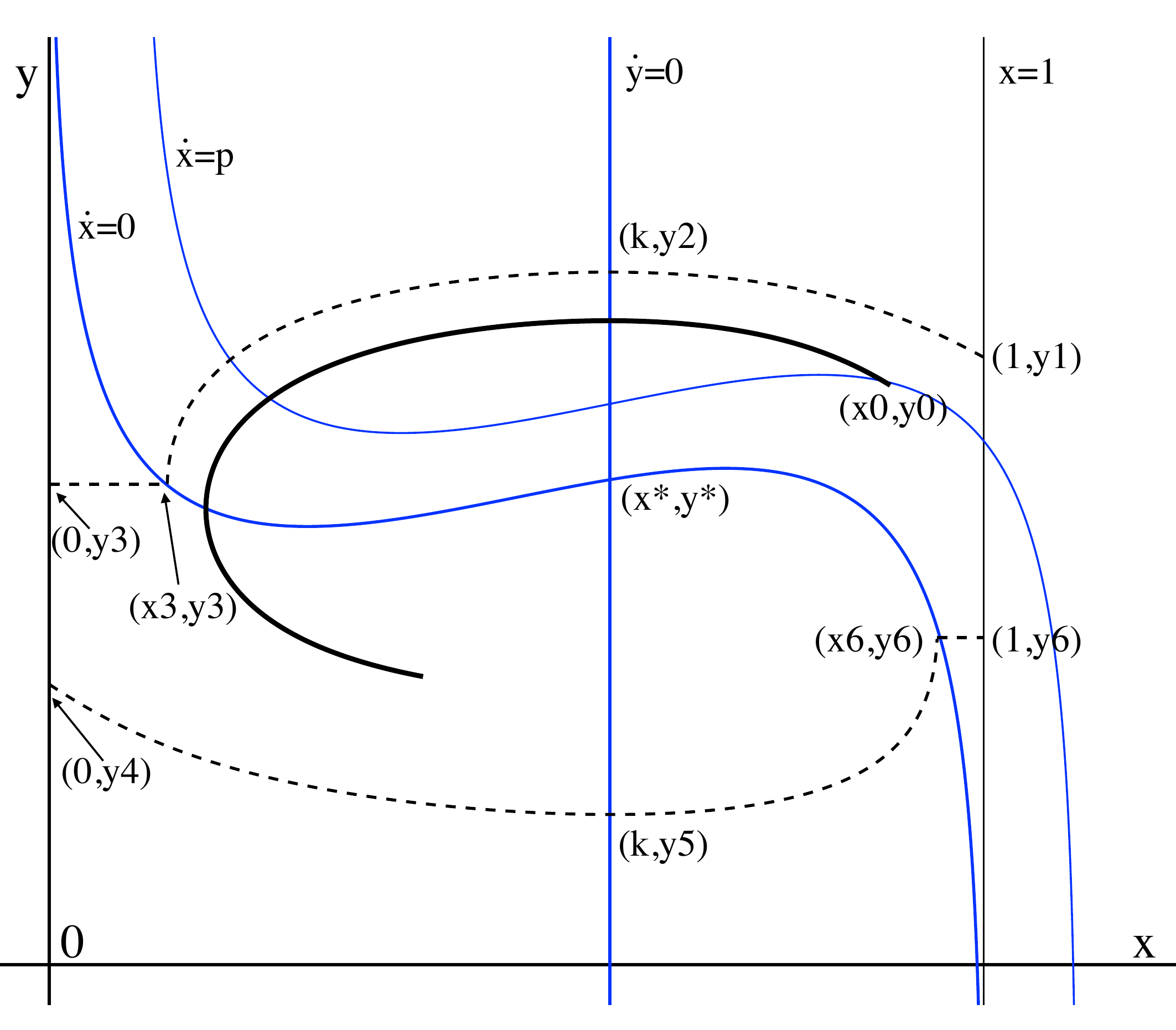}
\caption{Any trajectory (bold line) originating in the interior of the trapping region $E$ cannot cross its boundary (dashed lines). In this figure $k=0.6$, $r=0.25$, $s=1.25$, $\eta=0.25$, $p=-0.1$, and $x_0,y_0\approx (0.9, 0,.6)$.}  
\end{figure}

The solution from $(1, y_1)$ must flow northwest, so until it reaches the nullcline $\dot{y}=0$ it must remain in the region in which $\dot{x} < p$ and $\dot{y} >0$. Since the speed of $x$ is bounded away from 0, the $y$-nullcline $x=k$ will be reached in some finite time $T$ at some point $(k, y_2)$. But for $x \in [k, 1]$ the speed of $y$ is bounded from above: $\dot{y} = \frac{r}{s}(x-k) \le \frac{r}{s}(1-k)$, hence $y_2 \le y_0 + \frac{r}{s}(1-k)T$. Once the nullcline is reached, $y$ will have to decrease, so the solution trajectory from $(1, y_1)$ must lie below the line $y = y_2$ and ultimately intersects the x-nullcline at some $(x_3, y_3)$. Moreover, since the trajectory from $(x_0, y_0)$ (bold line in Figure 6) cannot intersect the trajectory from $(1, y_1)$, it must at some point escape into the region $B$ and subsequently escape from $B$ to $C$ through the segment of the x-nullcline connecting points $(x_3, y_3)$ and the steady state $(x^*, y^*)$. After it reaches $C$, we --following the same logic-- can find a point $(0, y_4)$ the trajectory from which (dashed curve from $(0, y_4)$ to $(x_6, y_6)$ in Figure 6) will have to intersect the $y$-nullcline at some point $(k, y_5)$, so that the solution from $(x_0, y_0)$ will have to transit back to region $A$ through region $D$ via the segment of the x-nullcline connecting points $(x_6, y_6)$ and the steady state $(x^*, y^*)$. Once it is back in $A$ it will again be bounded by the solution trajectory from $(1, y_1)$ and forced to complete another loop around the steady state. Therefore the solution trajectory from $(x_0, y_0)$ must remain within the closed region $E$ which boundary consists of the following curves: the solution trajectory from $(1, y_1)$ to $(x_3, y_3)$, the line segment from $(x_3, y_3)$ to $(0, y_3)$, the line segment from $(0, y_3)$ to $(0, y_4)$, the solution trajectory from $(0, y_4)$ to $(x_6, y_6)$, the line segment from $(x_6, y_6)$ to $(1, y_6)$, and the line segment from $(1, y_6)$ to $(1, y_1)$. In fact, any solution trajectory originating in the interior of $E$ cannot cross its boundary, since any such solution must flow northwest in $A$, southwest in $B$, southeast in $C$, and northeast in $D$ while not being able to intersect the trajectories from $(1, y_1)$ to $(x_3, y_3)$ and from $(0, y_4)$ to $(x_6, y_6)$. Since the steady state is repelling, there exists an open ball $B_{\epsilon}$ centered at the steady state such that all trajectories from $(x_0, y_0) \in E \setminus B_{\epsilon}$ are confined entirely to $E \setminus B_{\epsilon}$. Then since $E \setminus B_{\epsilon}$ is closed and bounded, positive invariant for the dynamics, and does not contain any steady states, any solution originating in that set must according to the Poincar\'{e}-Bendixson theorem converge to a closed orbit. 
\end{proof}
\bigskip
\begin{figure}[ht]
\centering
\includegraphics[scale=0.45]{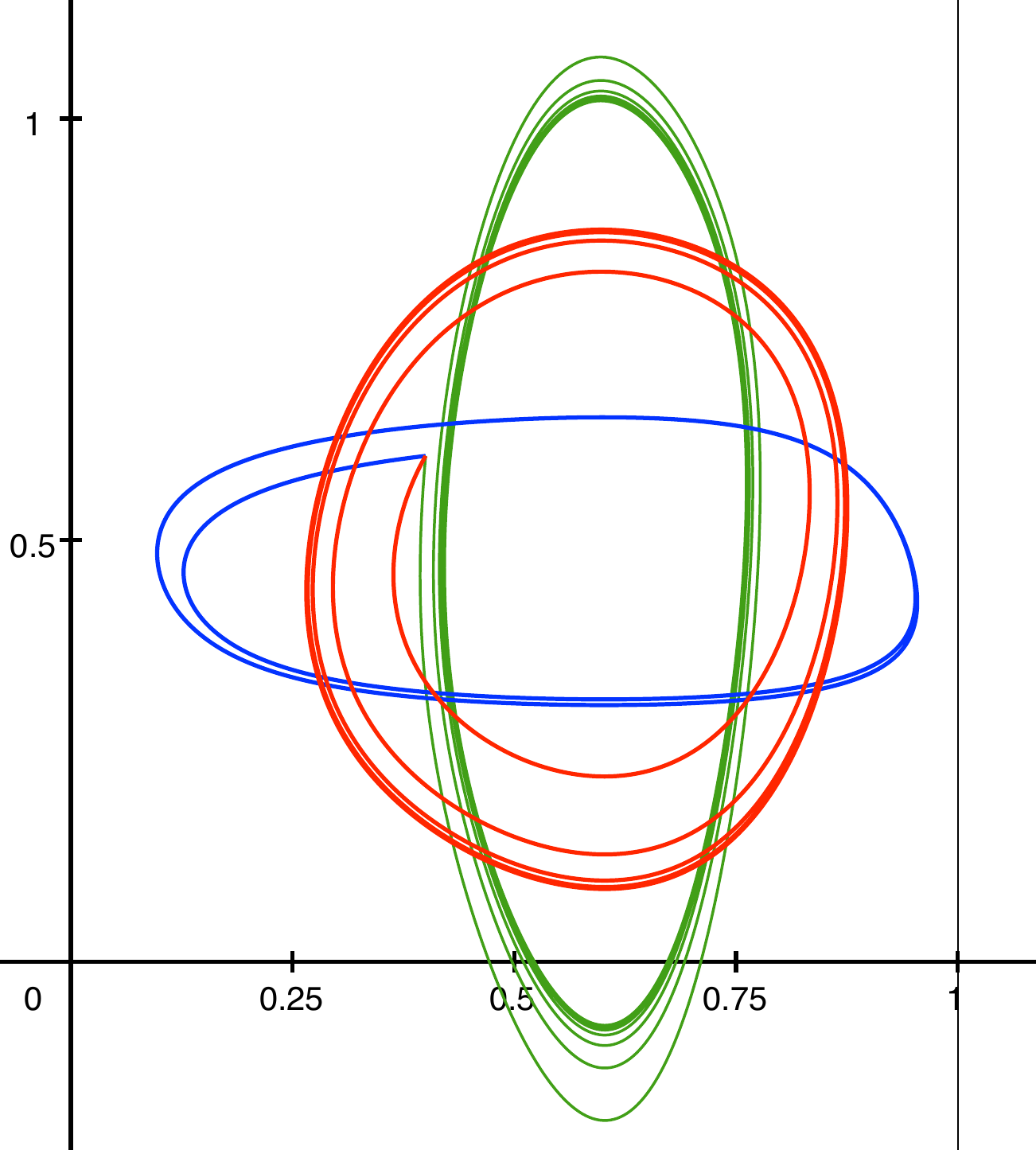}
\includegraphics[scale=0.45]{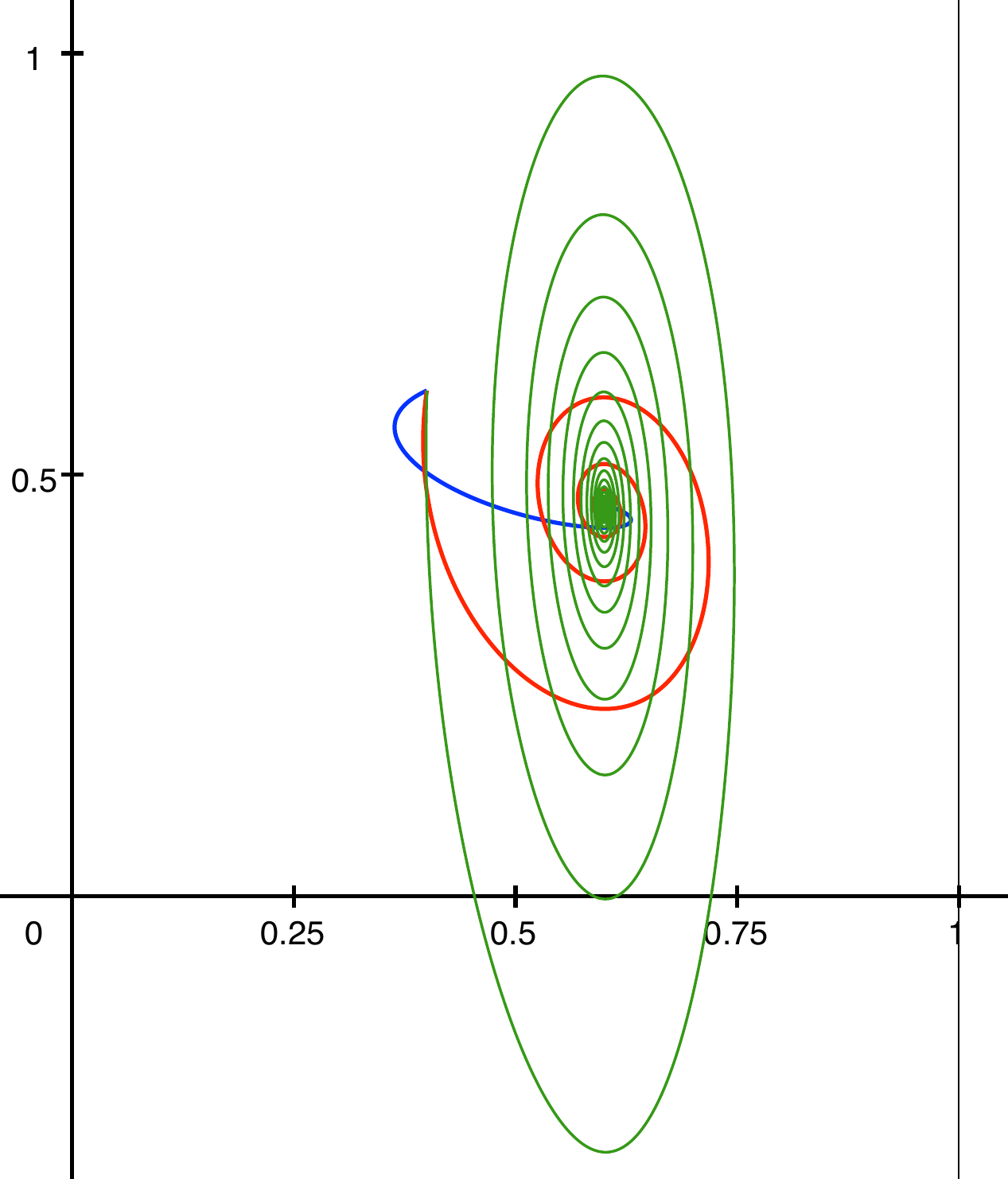}
\caption{Unstable ($\eta=\frac16$, left) and stable ($\eta=\frac13$, right) solution trajectories to the logit dynamic with parameters $k=0.6$ and $s=1$ from the initial condition $(x_0,y_0)=(0.4,0.6)$: $r=0.1$ in blue, $r=1.5$ in red, $r=10$ in green.}
\end{figure}

Based on the results of Propositions \ref{T1} and \ref{T2} we can see that the behavior of the system under the logit dynamic with small noise levels is very similar to that under the best response dynamic. In both cases the solution trajectories converge to orbits around the steady state, moreover as $\eta$ tends to 0 the steady state $(k, k-\frac{\eta}{s} \log{\frac{k}{1-k}})$ of the logit dynamic approaches the steady state $(k,k)$ of the best response dynamic. From multiple simulations (see examples in Figure 4 and Figure 7) we conjecture that a) for a fixed set of parameters solutions from all initial conditions converge in fact to the same orbit and b) as $\eta$ tends to 0 the limiting orbits of the logit dynamic converge to the orbit of the best response dynamic. In general it need not be the case that these two dynamics produce the same behavior -- see example A.1 in \cite{KojTak07}. Figure 7 also reinforces the observation about the role of the speed of the payoff change. As $r$ increases, the orbits shrink along the x-axis and stretch along the y-axis, as $y$ is able to adjust to changes in $x$ at a faster rate. 

\subsection{Replicator Dynamics}

The replicator dynamics, introduced in \cite{TayJon78}, emerges as the mean dynamics in populations in which agents are unable to determine the optimal strategy and use imitation to improve their performance\footnote{See \cite{BjoWei96} and \cite{Sch98}.}. As opposed to the best response and logit dynamics it does not require that the agents know the current population state. Instead it assumes that the share of the population playing a certain strategy grows at a rate proportional to the payoff advantage of that strategy. In two-strategy games this assumption implies that the best response and the replicator dynamic will have the same direction of motion, but the latter dynamic will be slower since an agent would switch only if he encounters someone who is already playing the optimal strategy. 

In terms of the parameters of the model the payoff advantage of strategy Left over the average payoff can be expressed as 
\begin{align*}
f_L(x) - \bar{f}(x) &= ax + b(1-x)- x(ax+b(1-x))-(1-x)(cx+d(1-x))=\\
&= x(1-x)(a-b-c+d)-(d-b)(1-x)=s(1-x)(x-y)
\end{align*}
and therefore we obtain the following system as the joint dynamics of payoffs and preferences:
\begin{align}
\dot{x} &=x\big(f_L(x)-\bar f(x)\big) = sx(1-x)(x-y) \label{eq11}\\
\dot{y} &=\frac{r}{s}(x-k) \tag{2}
\end{align}
We establish that, like in the case of the best response dynamic, the steady state of the replicator dynamic is repelling, but the behavior of the solution trajectories away from the steady state differs in these two cases.
\begin{prop}
Consider the joint dynamics generated by \eqref{eq11} and \eqref{pffadj2}. For any $r >0$\\
(i) the only steady state $(x^*,y^*) =(k,k)$ is repelling, \\ 
(ii) the solution trajectory from any initial condition spirals around the steady state and is unbounded.
\end{prop}
\begin{proof}
(i) For the state of indifference to be at rest, we must have $\dot{y} = 0$, so that $x=k$, but then $\dot{x}=0$ only if $y=k$, too. Therefore $(k,k)$ is the only steady state. We proceed by linearizing the system $F$ generated by \eqref{eq11} and \eqref{pffadj2} around the steady state. The Jacobian of $F$ is 
\begin{center}
$DF(x,y)=\left(\begin{matrix} s(2x-3x^2-y+2xy) & s(x^2-x)\\ \frac{r}{s} & 0 \end{matrix}\right)$  
\end{center}
Evaluated at the steady state $(k,k)$ it equals 
\begin{center}
$DF(k,k)=\left(\begin{matrix} sk(1-k) & -sk(1-k) \\ \frac{r}{s} & 0 \end{matrix}\right)$  
\end{center}
and generates the following characteristic polynomial:
\begin{align*}
\lambda^2- sk(1-k) \lambda + rk(1-k)=0
\end{align*}
Given that $s>0$, $r>0$, and $k \in (0,1)$ it must be that $sk(1-k) > 0$ and $rk(1-k) >0$. These conditions guarantee that the roots of the polynomial are either real and positive or complex with positive real part. Therefore the steady state is always repelling. \smallskip

(ii) Consider the function\footnote{I thank Matthew Johnston for discovering it.} $L: (0,1) \times \mathbb{R} \to \mathbb{R}$ 
\begin{align*}
L(x,y) = \frac{s}{2r}(y-k)^2 -\frac{k}{s} \log x - \frac{1-k}{s} \log (1-x)+ c_0  
\end{align*}
with $c_0 = \frac{k}{s} \log k + \frac{1-k}{s} \log (1-k)$. It has the following properties: \\
1) $L(x,y) \to \infty$ as $x \to 0$ or $x \to 1$ or $y \to \infty$, \\
2) $L(k,k) =0$. 

The critical points of $L$ are the solutions of the system
\begin{align*}
\frac{\partial L}{\partial x } =& -\frac{k}{sx} + \frac{1-k}{s(1-x)} = 0\\
\frac{\partial L}{\partial y } =& \frac{s}{r} (y-k) = 0
\end{align*}
Therefore $(x,y) = (k,k)$ is the only critical point of $L$. 

The Jacobian of $L$ at a point $(x,y)$ is 
\begin{center}
$DL(x,y)=\left(\begin{matrix} \frac{k}{sx^2} + \frac{1-k}{s(1-x)^2} & 0 \\ 0 & \frac{s}{r} \end{matrix}\right)$  
\end{center}
For all $(x, y) \in (0,1) \times \mathbb{R}$, $s, r >0$ and $k \in (0,1)$ it must be that 
\begin{align*}
\frac{k}{sx^2} + \frac{1-k}{s(1-x)^2} >0 \text{ and } \frac{s}{r} >0
\end{align*} 
so $DL(x,y)$ is positive definite and $L$ is convex. Therefore $(x,y) = (k,k)$ is the global minimizer of $L$, hence $L(x,y) >0$ for all $(x, y) \ne (k,k)$. 

Next observe that for $a>c_0$ the set $\{ (x,y)| L(x,y) \le a\}$ is bounded. To see this, consider the function $l(x,y) = (y-k)^2 - \log x - \log(1-x)$. All three terms of $l$ are non-negative, so $l(x,y) \le a$ implies $(y-k)^2 \le a$, $-\log x \le a$ and $- \log(1-x) \le a$, thus $k-\sqrt{a} \le y \le k+\sqrt{a}$ and $0 < e^{-a} \le x \le 1-e^{-a} <1$ (for the last inequality to hold, we must have $a \ge \log 2$). Hence any level set of $l$ (and thus of $L$) is bounded, and since $L$ is convex as the sum of three convex functions of one variable and a constant, the lower contour sets of $L$ must be convex as well, so that the level sets of $L$ have elliptical shape.  

Finally we show that $L(x,y)$ is nondecreasing along solutions of the system \eqref{eq11}-- \eqref{pffadj2}: 
\begin{align*}
\frac{dL}{dt} =& \frac{\partial L}{\partial x } \frac{dx}{dt}+\frac{\partial L}{\partial y }\frac{dy}{dt} = \\ 
=& \left(-\frac{k}{sx} + \frac{1-k}{s(1-x)} \right) sx(1-x)(x-y) + \frac{s}{r} (y-k) \frac{r}{s} (x-k) =\\
=& (x-k)(x-y) + (y-k)(x-k) = (x-k)^2 \ge 0
\end{align*}

By partitioning the state space into four regions using the nullclines of the dynamic it is straightforward to demonstrate that the solution trajectories must circle around the steady state (see the argument in part iii of Proposition \ref{T2}). Therefore, since the forward invariant set contains no steady states, any solution trajectory either approaches a closed orbit around the steady state or is unbounded. Our final step is to show that the former is never the case. 

Consider the trajectory from an initial condition $(x_0, y_0)$ with $L(x_0, y_0) = a >0$ and assume that it approaches a closed orbit, so that $\lim_{t \to \infty} L(x_t, y_t) = b < \infty$. Then the whole trajectory must be confined in a closed and bounded set $K = \{ (x,y) | a \le L(x, y) \le b\}$ which boundaries are the level sets $L(x,y) =a$ (inner ellipse in the Figure 8) and $L(x,y) = b$ (outer ellipse in the Figure 8). Fix a small $e>0$ and take the partition of $K$ with respect to the lines $x=k-e$ and $x=k+e$: $A = \{ (x,y) \in K | k+e \le x \le 1\}$, $B = \{ (x,y) \in K | k-e < x < k+e, x<y\}$, $C = \{ (x,y) \in K | 0 \le x \le k-e\}$, and $D = \{ (x,y) \in K | k-e < x < k+e, x>y\}$. 

\begin{figure}[ht]
\begin{center}
\includegraphics[scale=0.9]{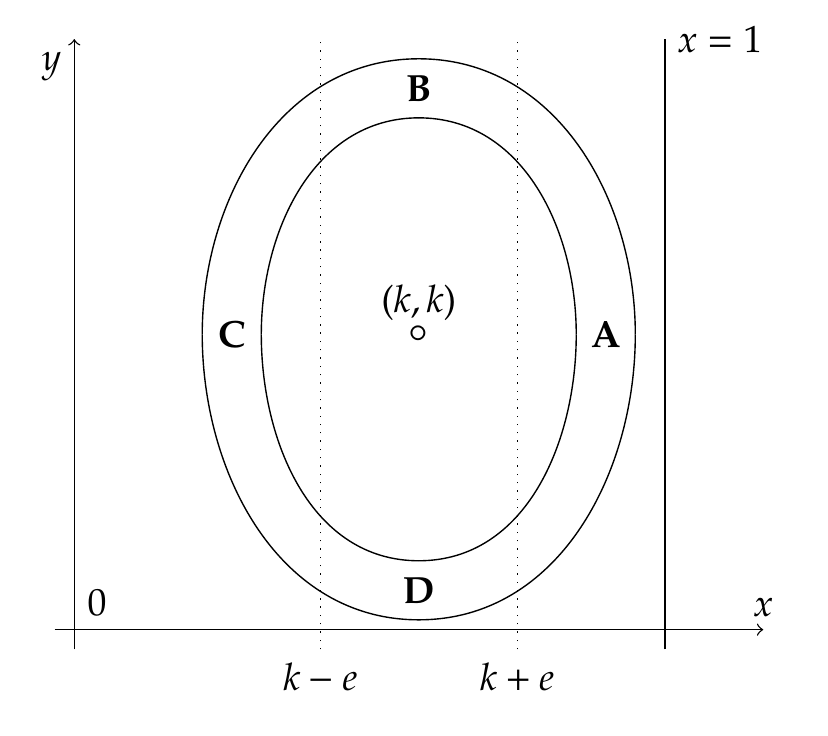}
\end{center}
\caption{The region $K$ between level sets $L(x,y)=a$ (inner ellipse) and $L(x,y)=b$ (outer ellipse) partitioned into four sets.}
\end{figure}

We know that the system would circle around the steady state visiting each of the regions $A$, $B$, $C$, and $D$ in turn, and our goal is to show that the time it takes to transit through $A$ or $C$ is bounded from below, whereas the time it takes to transit through $B$ or $D$ is bounded from above. If we choose $e$ to be small enough so that the diagonal does not intersect regions $B$ and $D$, the speed of $x$ in those regions will be bounded away from 0 and have constant sign. In $B$ with $x<y$ the speed of $x$ is negative: $\dot{x} = sx(1-x)(x-y) <0$, whereas in $D$ it would be positive. Let $v_B = \inf_{(x,y) \in cl(B)} |\dot{x}(x,y)|$ and $v_D = \inf_{(x,y) \in cl(D)} |\dot{x}(x,y)|$, where $cl(B)$ and $cl(D)$ are the closures of the corresponding sets. The existence of $v_B$ and $v_D$ is guaranteed by the fact that $\dot{x}(x,y)$ is continuous and $cl(B)$ and $cl(D)$ are compact. In addition, both $v_B$ and $v_D$ are positive. Then since $\dot{x}<0$ in $B$, the time it would take for the system to transit through that region is at most the time it will take for its $x$ component to travel from $k+e$ to $k-e$ with the lowest possible speed $v_B$. Thus it takes the system at most $\frac{2e}{v_B}$ to transit through $B$. Similarly, it takes at most $\frac{2e}{v_D}$ to transit through $D$.

Let $y_1, y_2$ be the $y$-components of the intersection of the line $x=k+e$ and the level set $L(x,y)=a$. Then they satisfy $L(k+e, y_1) = L(k+e, y_2)=a$. In the region $A$ the speed of $y$ is bounded and always positive: $\dot{y}(x,y) = \frac{r}{s}(x-k) >0$ and for $x \in [k+e, 1]$ we have $\dot{y}(x,y) \in [\frac{re}{s},\frac{r}{s}(1-k)]$, so it takes at least $t_A=\frac{|y_1-y_2|}{\frac{r}{s}(1-k)}$ to transit through $A$. Similarly, it takes at least $t_C=\frac{|y_3-y_4|}{\frac{r}{s}k}$ where $y_3, y_4$ satisfy $L(k-e, y_3) = L(k-e, y_4)=a$ to transit through the region $C$. 

Finally, observe that for $(x,y) \in A \cup C$ the time derivative of the function $L$ along solutions of the system is bounded from below: $\frac{dL}{dt}(x,y) = (x-k)^2 \ge e^2$. Let $T>0$ and $m_A$ and $m_C$ be the number of times the solution trajectory from $(x_0, y_0)$ transited through the regions $A$ and $C$, correspondingly, during the time span $[0, T]$. Then 
\begin{align*}
L(x_T, y_T) = L(x_0, y_0) + \int_{0}^{T} \frac{dL}{ds}(x_s,y_s) ds \ge L(x_0, y_0) + m_A t_A e^2 + m_C t_C e^2
\end{align*}
By assumption, $L(x_T, y_T) \le b$, but on the other hand as $T \to \infty$, both $m_A \to \infty$ and $m_C \to \infty$, therefore it must be that $L(x_T, y_T) \to \infty$, too. Intuitively, every time the solution trajectory passes through region $A$ the value of $L$ must grow by at least $t_A e^2$, and since the solution passes through that region infinitely many times, the value of $L$ must grow unboundedly. Therefore all trajectories of the system must form unbounded spirals around the steady state.
\end{proof}
\bigskip 

Unlike in the best response case the trajectories of the replicator dynamic are unbounded and form unstable spirals around the steady state (see Figure 9). 

\begin{figure}[ht]
\centering
\includegraphics[scale=0.4]{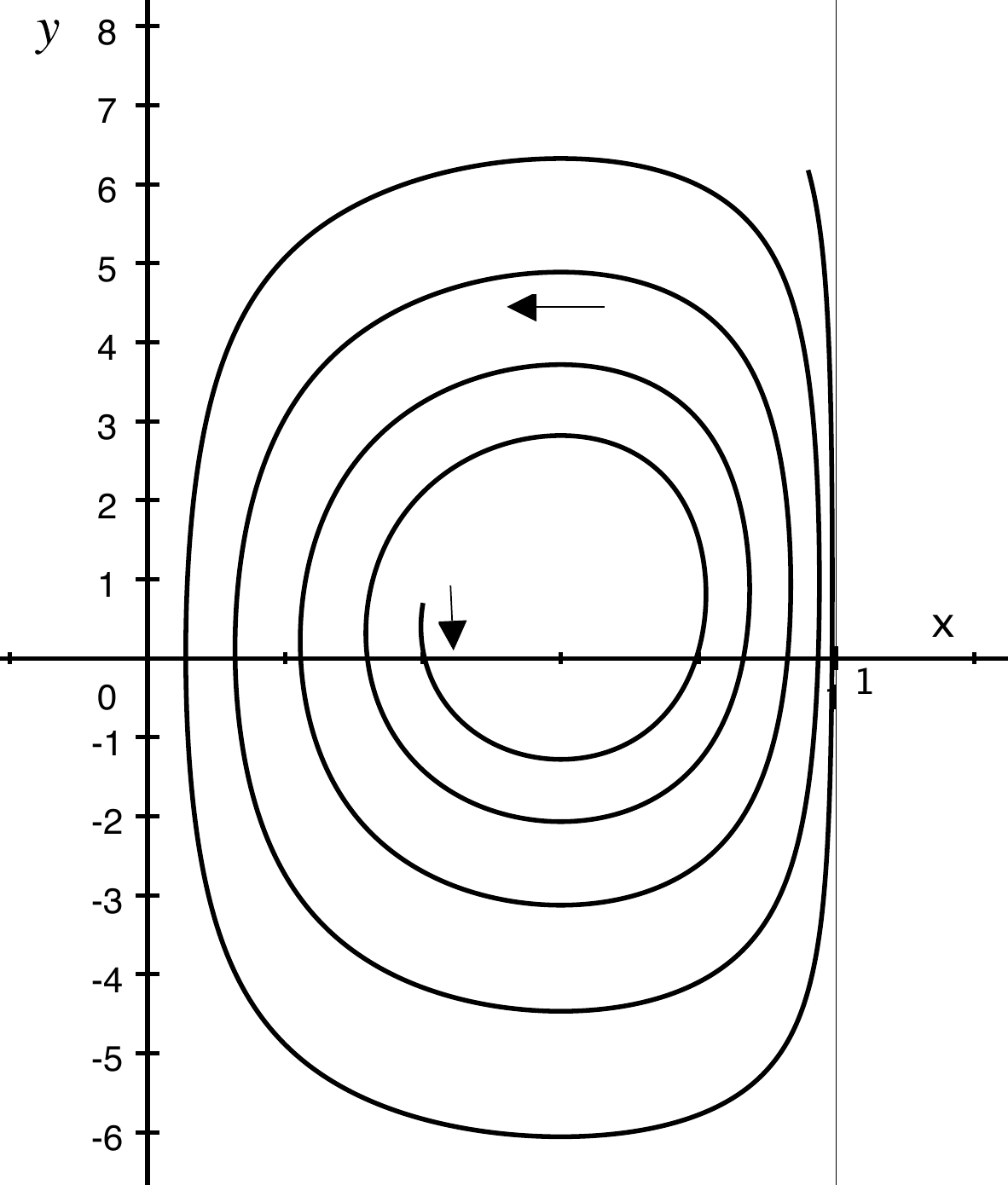}
\caption{Solution to the replicator dynamic with parameters $k=0.6$, $s=0.17$, $r=3$ from initial conditions $(0.4, 0.7)$.}
\end{figure}

This difference should be attributed to the fact that the law of motion of the population state depends on the state of indifference in the replicator case (equation \eqref{eq11}), whereas in the best response case it does not (equation \eqref{BR}). Intuitively, in the latter case the rate at which the agents switch to the current best response does not depend on the magnitude of the payoff advantage of that strategy whereas under imitative behavior the switch rate is the higher the higher the payoff difference. Thus the replicator dynamic is capable of 'accumulating' the payoff advantage for a given population state, so that after each cycle of the system the switch rate at a given population state increases, and so does the share of states at which the base game has a dominant strategy.

\section{Conclusion} \label{c1}
In this paper we introduced an evolutionary model in which the aggregate behavior of the population affected future individual preferences by influencing the payoffs of the underlying two-strategy game. The payoffs to strategies decreased at rates proportional to the intensity of their use, thus equilibrating any advantage that one strategy would have over the other. In a homogeneous population of myopic individuals this process gave rise to cyclical behavior as the agents would from time to time switch the strategy on which they were trying to coordinate.

We derived the joint dynamics of the aggregate behavior of the population and the individual preferences and analyzed the long-term behavior of the resulting system. We demonstrated that under the best response dynamic the system admits a unique steady state and that solution trajectories from all initial conditions other than the steady state converge to orbits around it. This result extends to the logit dynamic with small noise levels, while at large noise levels the steady state becomes a sink. Under the replicator dynamic the unique steady state is repelling and the solution trajectories form unstable spirals around it. 

There are several directions in which these results can be extended. First, there are some technical questions that remain unanswered: we conjectured that the limiting behavior of the best response and the logit dynamics with small noise levels is independent of the initial conditions, that is, for a given dynamics all solutions converge to one and the same closed orbit. A related question is whether, as noise level vanishes, the limiting behavior of the logit dynamics approaches that of the best response. 

Second, one can possibly generalize our results to the class of payoff-monotonic dynamics, and consider more general forms of the payoff adjustment function.

Third, it remains unclear whether cyclical behavior may survive in our model under the perfect foresight dynamics of \cite{MatMat95}. As \citet{Rapp} demonstrates for 'bad' Rock-Paper-Scissors games, the solution trajectories under the PFD are still cycles, albeit different from the myopic best-response cycles. His intuition seems to apply to our model as well, since the equilibrating effect guarantees no strategy can remain the unique best response indefinitely. Thus along any perfect foresight equilibrium path it must be either that both strategies produce the same payoffs or that the best response strategies alternate. A formal analysis of perfect foresight dynamics in our environment is an interesting question for future research. 

Finally, the perfect foresight dynamics can help view our model in the light of differential games literature. \cite{HofSor02} construct an $N$-player differential game that mimics the interaction between $N$ populations of agents under perfect foresight, and demonstrate that its equilibria are in one-to-one correspondence with the equilibrium paths of the population game. Since in our model the interaction takes place within a single population, the attention in the corresponding differential game should be restricted to symmetric equilibria only. 
\end{singlespace}
\bibliographystyle{apalike}
\bibliography{thesis}
\end{document}